\documentclass[prc,aps,floatfix,nofootinbib,preprint]{revtex4} 
\usepackage{graphicx} 
\usepackage{amssymb} 
\usepackage{amsmath} 
\usepackage{amsfonts}

\def\be{\begin{equation}} 
\def\ee{\end{equation}}

\def\ev8{{\tt ev8}} 
 
\begin{document} 
\title{Derivation of $K$-matrix reaction theory in a discrete basis
formalism
} 
\author{Y. Alhassid,$^{1}$ G.F.~Bertsch,$^{2}$ and P. Fanto$^{1}$ } 
\affiliation{$^{1}$Center for Theoretical Physics, Sloane Physics 
Laboratory, Yale University, New Haven, Connecticut 06520, USA\\ 
$^{2}$Department of Physics and Institute for Nuclear Theory, 
Box 351560\\ University of Washington, Seattle, Washington 98195, USA} 
 
\begin{abstract} 
 
The usual derivations of the $S$ and $K$ matrices for two-particle reactions 
 proceed through the Lippmann-Schwinger equation with 
formal definitions of the incoming and outgoing scattering states.
Here we present a simpler alternative derivation 
that is carried out completely in the Hamiltonian representation, using a
discrete basis of configurations for the scattering channels as well
as the quasi-bound configurations of the combined fragments.  
We use matrix algebra to derive an explicit expression  
for the $K$ matrix in terms of the Hamiltonian of the internal states of the compound system and the coupling 
between the channels and the internal states.  
The formula for the $K$ matrix includes explicitly a real dispersive shift matrix to the internal Hamiltonian that is easily
computed in the formalism. That expression is applied to derive the usual
form of the $S$ matrix as a sum over poles in the complex energy plane.   
 Some extensions and limitations of the discrete-basis Hamiltonian formalism are discussed in the concluding remarks 
and in the Appendix.

\end{abstract} 
 
\maketitle 
 
\section{Introduction} 
 
The $K$-matrix formalism for reactions between 
particles with an internal structure is widely used in many domains of physics, including
molecular collisions~\cite{lin19}, mesoscopic physics~\cite{fyo96,alh00}, 
hadronic spectra~\cite{lon86}, nuclear reactions~\cite{kaw15}, and statistical 
reaction theory in general~\cite{mit10}.  
Its advantage over the competing $R$-matrix theory~\cite{wig47,lane58, desc10} is a
simplified connection between internal states of the compound
system and the channel wave functions of the incoming or outgoing
particles.\footnote{We use the term `particles' both for the elementary
constituents and the (possibly composite) reactants 
in the initial or final states of the reaction.}
In particular, in the $K$-matrix approach, the Hamiltonian dynamics within the internal
states can be treated by well-known configuration-interaction (CI)
methods~\cite{low55}.  However, the derivation of equations relating the $K$ matrix
to the Hamiltonian can be rather obscure
in the literature. The derivations often start from the Lippmann-Schwinger
equation and its associated $T$ matrix, which is already several steps removed from the Hamiltonian
equation expressed in a computationally transparent basis~\cite{dal61,mah69,chu95,tay06,mit10}.  Here
we carry out the derivations starting from a representation of the Hamiltonian
$H$ in a discrete basis.
As a benefit, we find an expression for the dispersive couplings
of the internal states to the continuum that is computationally quite
simple.  In contrast, many derivations in the literature suppress
these terms in the final formulas.  
 
  A simple version of our formalism has been applied in
nuclear reaction theory~\cite{fan18,ber18}.  In the Mazama code
introduced in Ref.~\cite{fan18}, the diagonal $S$-matrix element is computed for
one specific channel, providing the elastic cross section in that channel and the total reaction cross section.
 Here we consider a general scattering problem of any number of two-particle channels.

\section{Discrete-basis formulation of the scattering problem}

\subsection{Discretized two-particle Hilbert space}

The Hilbert space of the two-particle scattering system consists of two subspaces.  
The first contains configurations, labeled by $\lambda$, that are used to
construct internal wave functions of the compound system; 
the scattering wave function amplitude for each internal configuration $\lambda$ will be denoted $\psi_\lambda$.
The second subspace contains all 
the scattering channels.  Each channel $c$ is defined by
the set of configurations having the same internal structures for the two
particles and differing only in the relative coordinate between the particles'
centers of mass. 
We introduce a discretized mesh of separation distances $r_n = R_0 + n \Delta r$ ($n=0,1,\ldots$) 
with finite spacing $\Delta r$.
The channel wave function in channel $c$ then consists of the
set of amplitudes $\varphi_c(n)$ of the configurations on the mesh points,
\be
\label{phi_c}
\vec \varphi_c  = \{ \varphi_c(0),\varphi_c(1),...\} \;.
\ee
 $R_0$ is assumed to be sufficiently large such that 
potential interactions between the reactants at larger 
distances $r >R_0$ can be ignored.  The first configuration $\varphi_c(1)$   
is connecting to the internal states, either
directly or through some extension of the chain into the interacting
region.  A less restrictive definition of the channel wave
function that allows for a potential interaction $V_c(r)$  in each channel $c$ is given in Appendix A.
 
\subsection{Hamiltonian matrix elements}

\subsubsection{Channel Hamiltonian}

The Hamiltonian in the channel space is taken to be
the kinetic energy operator of the relative motion of the two particles.  It is approximated by the second-order
difference formula on neighboring mesh points.  
Following nomenclature from condensed-matter physics, we denote the
Hamiltonian matrix element between adjacent states in the
channel by $t_c$ (here $t_c=\hbar^2/2 M_c(\Delta r)^2$ where $M_c$ is the reduced mass of the two fragments).
Then the Hamiltonian matrix $H^c$ describing the relative motion of the fragments in channel $c$ has the matrix elements
\be
\label{H^c}
H^c_{n,n'} = -t_c \delta_{n,n'+1} + (2 t_c +E_c ) \delta_{n,n'}
-t_c \delta_{n,n^\prime-1} \;,
\ee
where $E_c$ is the summed energy of the two reactants at rest.
In the region $r_n > R_0$ the Hamiltonian is invariant under translations, 
so its eigenfunctions at energy $E$ can be expressed as a superposition of an incoming wave and an outgoing wave with
 wave number $k_c$ and amplitudes $a_c^{(-)}$ and $a_c^{(-)}$, respectively
\be
\label{a_c}
\varphi_c(n) = a_c^{(-)} e^{-ik_c r_n} - a_c^{(+)} e^{i k_c r_n } \;.
\ee
Using $[H^c \vec \varphi_c](n) = E \varphi_c(n)$ for $n >0$ together with (\ref{a_c}), the energy-momentum dispersion  is given by 
\be\label{k_c}
 E-E_c =2t_c(1-\cos \kappa_c) \;.
\ee
where 
 $\kappa_c=k_c \Delta r$. In the continuum limit, Eq.~(\ref{k_c}) reduces to the usual quadratic dispersion
  $E-E_c= (\hbar^2/ 2 M_c) k_c^2$.
  
\subsubsection{Interaction with internal states}

\begin{figure}[tb] 
\begin{center} 
\includegraphics[width=0.8\columnwidth]{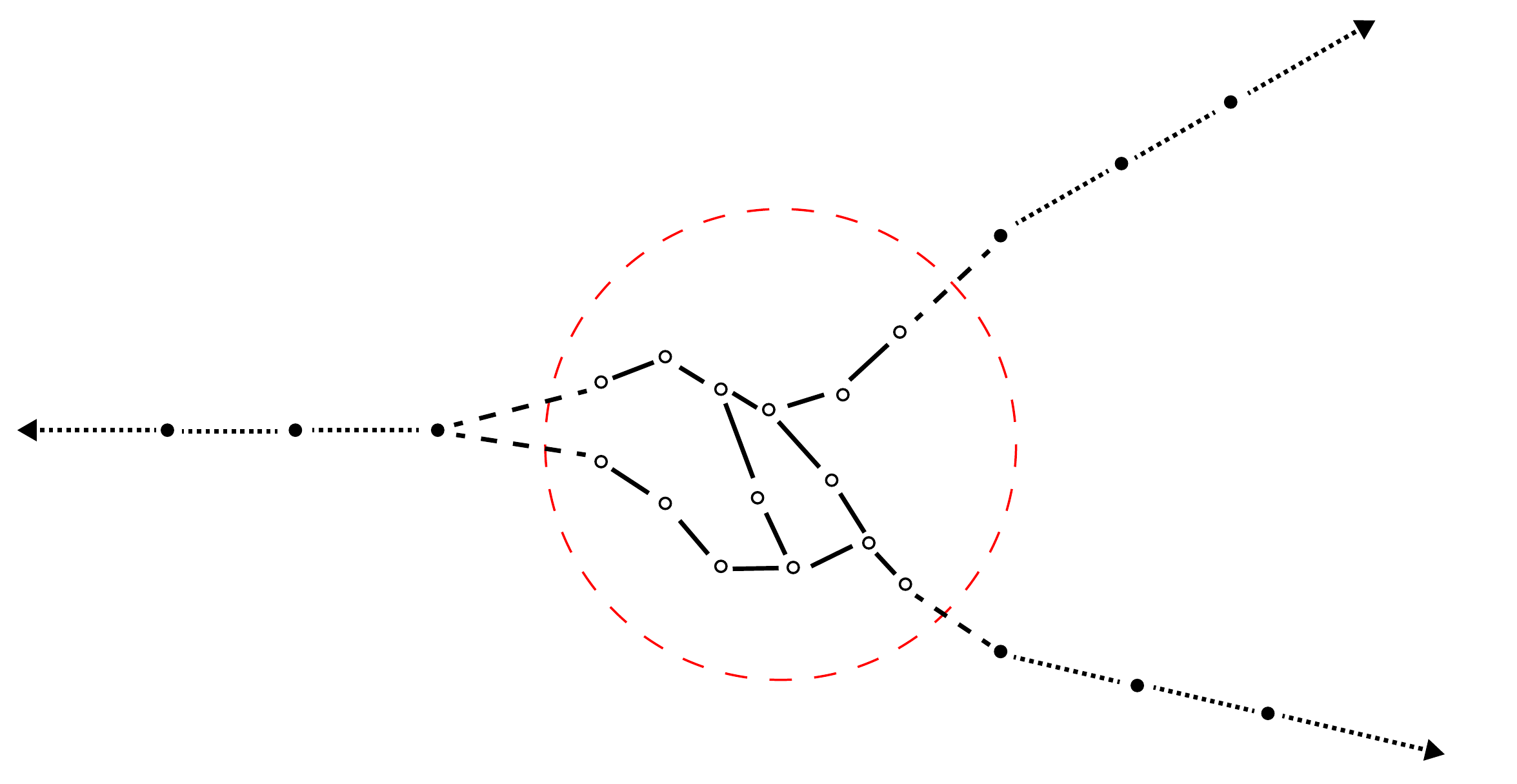} 
\caption{Connectivity of the discretized Hamiltonian.  The
internal states are enclosed in the large dashed circle. 
Small open circles represent states
of the internal Hamiltonian and the solid lines indicate off-diagonal
matrix elements of the internal Hamiltonian.
Solid circles represent
the discretized channel configurations.  They are coupled to each other
through the dotted lines to generate the channel Hamiltonian.    
The dashed lines denote matrix elements $v_{\lambda,c}$ connecting the channels to the internal states.
\label{Hconnectivity}
} 
\end{center} 
\end{figure}

The Hamiltonian matrix elements involving states in  the interaction
region $r \leq R_0$ are of 
two kinds: those strictly between internal states and those
that connect with the channel wave functions at the $n=1$ site.
We denote the latter matrix elements connecting the internal state $\lambda$ with channel $c$ by  $v_{\lambda,c}$.  
Fig.~\ref{Hconnectivity} demonstrates the states and the Hamiltonian matrix elements
that connect them.

We consider $N_i$ internal states and $N_c$ channels, and assume that the internal state Hamiltonian is diagonal
 with energies $E_\lambda$.  For each channel $c$, the wave function  is regular at $n=0$, i.e., $\varphi_c(0)=0$.
  At radial site $n = 1$ the scattering wave function satisfies the Hamiltonian equation 
\be
\label{H1}
-t_c \varphi_c(2) + ( 2 t_c+E_c) \varphi_c(1) +  \sum_{\lambda = 1}^{N_{i}} v_{\lambda,c} \psi_\lambda  =  E  \varphi_c(1) \,, \qquad c = 1,...,N_c \;,
\ee 
while the corresponding equations for the internal-state amplitudes are
\be
\label{Hr}
\sum_{c=1}^{N_c} v_{\lambda,c}\varphi_c(1) + E_\lambda \psi_\lambda =  E \psi_\lambda \,,\qquad \lambda = 1,...,N_i \;.
\ee
Eliminating the internal state amplitudes $\psi_\lambda$  from Eqs.~(\ref{Hr}) and substituting in Eqs.~(\ref{H1}), we find
\be\label{Hcc'}
-t_c\varphi_c(2) + (2t_c \cos\kappa_c) \varphi_c (1) + \sum_{\lambda, c'} \frac{v_{\lambda c} v_{\lambda c'}}{E-E_\lambda} \varphi_{c'}(1) =0\;.
\ee

\section{$D$ matrix} 

Substituting the channel wave function form (\ref{a_c}) in Eqs.~(\ref{Hcc'}) yields a set of coupled linear equations relating the 
vector of outgoing amplitudes $\vec a^{(+)}= (a_1^{(+)},a_2^{(+)},...,a_{N_c}^{(+)})$ 
to the vector of incoming amplitudes
$\vec a^{(-)}= (a_1^{(-)},a_2^{(-)},...,a_{N_c}^{(-)})$
\be\label{a-a+}
t_c a_c^{(-)} + \sum_{\lambda c'} \frac{v_{\lambda c} v_{\lambda c'}}{E-E_\lambda} e^{-i\kappa_{c'} }a_{c'}^{(-)} =
 t_c a_c^{(+)} + \sum_{\lambda c'} \frac{v_{\lambda c} v_{\lambda c'}}{E-E_\lambda} e^{i\kappa_{c'} }a_{c'}^{(+)} \;,
\ee 
where we have absorbed a factor of $e^{-ik_c R_0}$ in $ a_c^{(-)}$ and a factor of $e^{ik_c R_0}$ in $ a_c^{(+)}$.

In principle, we could define an $N_c \times N_c$ matrix
that transforms $\vec a^{(-)}$ to $\vec a ^{(+)}$ but this is 
 not the $S$-matrix.  The $S$-matrix preserves the total 
 probability flux and requires the amplitudes
$a_c^{(\pm)}$ to be normalized to the unit flux.  To change to
flux-normalized variables, we note that, for a tridiagonal channel Hamiltonian, the probability current $J_c(n \to n+1)$  from a
site $n$ to the neighboring site $n+1$ is given by
\be\label{current}
J_c(n \to n+1) = i H^c_{n,n+1} [\varphi_c(n)\varphi_c^*(n+1) - \varphi^*_c(n)\varphi_c(n+1))]\;
\ee
up to a channel-independent constant. 
Applying (\ref{current}) to the wave functions $a_c^{(\pm)} e^{\pm i k_c r_n}$ for the Hamiltonian (\ref{H^c}), we find for the current
$J_c$ in channel $c$
\be\label{model_curr}
J_c = \pm 2 t_c \sin \kappa_c \, |a_c^{(\pm)}|^2 \;,
\ee
which is independent of $n$.
The flux-normalized amplitudes are thus
\be\label{bc_def}
b_c^{(\pm)} = a_c^{(\pm)}/d_c \;,
\ee
where
\be\label{dcdef}
d_c = \left(2 t_c \sin \kappa_c\right)^{-1/2} \;.
\ee

Eqs.~(\ref{a-a+}) can be rewritten for the flux-normalized amplitudes $b_c^{(\pm)}$
\be\label{b-b+}
d_c t_c b_c^{(-)} + \sum_{\lambda c'} \frac{v_{\lambda c} v_{\lambda c'}}{E-E_\lambda} d_{c'}e^{-i\kappa_{c'} }b_{c'}^{(-)} =
d_c t_c b_c^{(+)} + \sum_{\lambda c'} \frac{v_{\lambda c} v_{\lambda c'}}{E-E_\lambda} d_{c'} e^{i\kappa_{c'} }b_{c'}^{(+)} \;.
\ee
Dividing both sides of the equation by $d_c t_c$, we obtain
\be\label{DbDb}
D \vec b^{(-)} = D^* \vec b^{(+)} \;,
\ee
where the matrix $D$ is defined by\footnote{The definitions include factors
of $2\pi$ and $(2\pi)^{-1/2}$ following the convention in the
literature~\cite{mit10}}.
\be\label{Ddef}
D_{c,c^\prime} =  \delta_{c,c^\prime} + 2\pi \sum_{\lambda=1}^{N_i} \frac{W_{\lambda c}  W_{\lambda,c^\prime}}{E - E_\lambda} d^2_{c^\prime} t_{c'} e^{-i\kappa_{c'}} 
\ee
with
\be\label{Wdef}
W_{\lambda c}  =\frac{1}{\sqrt{2\pi}}\frac{v_{\lambda c} } {d_c t_c} \;.
\ee
$D^*$ is obtained from $D$ by simply replacing $e^{-i\kappa_{c'}} \to e^{i \kappa_{c'}}$. 
The $S$ matrix is defined by $ \vec b^{(+)}  = S \vec b^{(-)}$ and, using Eq.~(\ref{DbDb}), is given by
\be
\label{SDD}
S={D^*}^{-1} D \;.
\ee

\section{$K$ matrix}

The $K$ matrix is defined from the $S$ matrix by the implicit relation
\be
\label{S-K}
S = \frac{1+iK}{1-iK} \;.
\ee
Substituting Eq.~(\ref{SDD}) in Eq.~(\ref{S-K}), and solving for $K$, we express $K$ in terms of $D$ and $D^*$
\be\label{KfromD}
K =- i (D+D^*)^{-1} (D-D^*)\;.
\ee

In the following we derive an explicit expression for the matrix elements of  $K$.  Using Eq.~(\ref{Ddef}), we have
\be\label{DpmD*} 
\frac{D + D^*}{2} =  1+ W^T (E - H)^{-1} V \;,\;\;\; \frac{D - D^*}{2} = -i \pi W^T (E - H)^{-1} W \;,
\ee
where 
\be
H=\sum_\lambda | \lambda\rangle E_\lambda \langle \lambda| 
\ee
is the internal state Hamiltonian of the compound system. The matrix $V$ is defined by
\be\label{V}
V_{\lambda c} = \pi W_{\lambda c} \cot \kappa_c \;,
\ee
where we have used $d_c^2 t_c =(2 \sin \kappa_c)^{-1}$.

Substituting Eq.~(\ref{DpmD*}) in (\ref{KfromD}), we  find
\be\label{K-X}
K = -(1+X)^{-1} X \tan \kappa = -[1-(1+X)^{-1}]  \tan \kappa \;,
\ee
where the matrix $X$ is defined by
\be
X= W^T (E - H)^{-1} V\;,
\ee
and $\tan \kappa$ is a diagonal matrix with elements $\tan \kappa_c$ along its diagonal.

To invert  $1+X$ we use the operator identity $B^{-1}(B-A)A^{-1} = A^{-1}-B^{-1}$ with $A=E- H$ and $B=E-H +V W^T$ to find
\be
(E-H +V W^T)^{-1} V W^T (E-H)^{-1} = (E-H)^{-1} - (E-H + V W^T)^{-1} \;.
\ee
Multiplying by $W^T$ on the left and by $V$ on the right, we obtain
\be\label{X-Y}
YX=X-Y \;,
\ee
where 
\be
Y=  W^T (E-H+ VW^T )^{-1} V \;.
\ee
Solving (\ref{X-Y}), we find $(1+X)^{-1}=1-Y$. Substituting in (\ref{K-X}), we find
\be
K=- Y \tan\kappa = -W^T (E-H+ VW^T)^{-1} V \tan\kappa =  -\pi W^T (E-H+ VW^T)^{-1} W \;,
\ee
where we have used Eq.~(\ref{V}).

 The final expression for $K$ is thus
\be\label{K-matrix}
K=  \pi W^T (H+\Delta -E)^{-1} W \;,
\ee
where  $W$ is given in Eq.~(\ref{Wdef}) and describes the coupling matrix of the channels to the internal states, while $\Delta = - VW^T$ is the real shift matrix 
\be\label{Delta_def}
\Delta_{\lambda \lambda'}  =   -\pi \sum_c W_{\lambda c}  W_{\lambda' c}  \cot \kappa_c \;.
\ee

The above expression for $K$ has the same form as the usual $K$ matrix, c.f.~Eq.~(18) of Ref.~\cite{alh00}. 
However, our term includes the real shift
matrix $\Delta$ that is usually ignored in expressions for the $K$ matrix. 
In other derivations of the $S$ matrix, this shift arises from off-shell
couplings to the channels; see, e.g., Eqs.~(28-30) of Ref.~\cite{mit10}.  In
our approach, this shift arises naturally from the matrix algebra.

The $K$ matrix in (\ref{K-matrix}) is real symmetric, which guarantees that the $S$ matrix in (\ref{S-K}) is symmetric and unitary. 

\section{$S$ matrix}

To find an explicit expression for the $S$ matrix, we use again the operator identity ${B^{-1}(B-A)A^{-1} = A^{-1}-B^{-1}}$ but now for $A=E- (H+\Delta)$ and
 $B=E- (H+\Delta -i\pi WW^T)$. We obtain
\begin{eqnarray}
i \pi [E- (H+\Delta -i\pi WW^T)]^{-1} W W^T [E-(H+\Delta)]^{-1}  \nonumber \\
= [E- (H+\Delta)]^{-1} - [E- (H+\Delta -i\pi WW^T)]^{-1} \;.
\end{eqnarray}
Multiplying by $\pi W^T$ on the left and by $W$ on the right, we find
\be \label{K-Z1}
i\pi ZK = K+\pi Z  \;,
\ee
where
\be\label{Z}
Z = W^T [E- (H+\Delta -i\pi WW^T)]^{-1}  W \;,
\ee
and we have used the expression (\ref{K-matrix}) for the $K$ matrix.  Relation (\ref{K-Z1}) can be rewritten in the form
\be\label{K-Z2}
\frac{K}{1-iK} = -\pi Z \;.
\ee
 Using the relation (\ref{S-K}) between the $S$ matrix and the $K$ matrix, we find
 \be
 S=1+2i \frac{K}{1-iK} = 1 - 2\pi i Z
 \ee 
where we have used (\ref{K-Z2}) to obtain second equality.  We thus find an explicit expression for $S$
\be
S= 1 - 2\pi i W^T [E- (H+\Delta -i\pi WW^T)]^{-1}  W \;,
\ee
which includes both a real shift $\Delta$ and an imaginary shift $-i\pi WW^T$ to the Hamiltonian $H$. This expression coincides formally 
with Eqs.~(28-30) in Ref.~\cite{mit10} for the $S$ matrix in the absence of background scattering. 
 
\section{Concluding remarks}

We have described an alternative derivation of the $K$ matrix of
scattering theory, as shown for the Hamiltonian specified
by Eqs.~(\ref{H^c}), (\ref{H1}) and (\ref{Hr}).  Using this derivation, we are able to avoid imposing
formal structures such as the continuum Green's functions of
Lippmann-Schwinger reaction theory.  It practice, it is well-suited to
many-body Hamiltonians of equal-mass particles, in which case it may be difficult to 
identify a relative coordinate.  This includes nuclei and atomic
condensates where common practice follows the
Hartree-Fock or Hartree-Fock-Bogoliubov approximations and their extensions
in the CI framework.  For large systems, this approach needs much less 
computational effort than other reaction formalisms, which rely on explicit
antisymmetrization and/or the use of a Jacobi coordinate representation
to separate out a channel wave function $\varphi^r_c(r)$ in the relative
coordinate $r$ of the two particles.

In the above derivation, we left unspecified the exact relationship of the
usual channel wave function $\varphi^r_c$ to the discrete-basis wave function 
$\vec \varphi_c$.  These quantities have different
dimensions:  the components of $\vec \varphi_c$ are dimensionless amplitudes in the CI formalism
while $\varphi^r_c$ has dimension $[{\rm length}]^{-1/2}$, the same as ordinary
coordinate-space wave functions.  The formal connection between the two
is not obvious, since it is difficult to separate out a relative
coordinate wave function unless it is already defined in the CI basis.
Our approach only involves the role of the relative coordinate at large
separations, where the absence of interactions leads to 
the simplified Hamiltonian approximation in Eq.~(\ref{H^c}). 

The present formalism might be  applicable
to problems in nuclear reaction theory such as fission~\cite{ber19}.  It should also 
 simplify the treatment of the interaction between droplets
of atomic condensates, such as the fusion reaction described in Ref.~\cite{shin04}. 

\section*{Acknowledgements} 

We thank J.J.~Rehr for discussion on possible applications to molecular
reactions.  The work of Y.A.  and P.F.  was supported in part by the U.S. DOE grant No.~DE-SC0019521, and by the U.S. DOE NNSA Stewardship
Science Graduate Fellowship under cooperative agreement No.~DE-NA0003864. 
 
\appendix 
\section{Potential interactions in the channels}

Our definition of the channel Hamiltonian requires that the 
starting point at $R_0$ be beyond the range of the interaction in the given channel.  This
is obviously inefficient if there are 
long-range potential interactions between the reactants.  As in other formulations
of reaction theory, the present framework can include
elastic scattering potentials $V_c(r)$ in the channels to reduce the size of the
interaction zone.

We define a mesh for which $n=1$ is the point where the channel configurations interact
with the internal ones. First we solve the one-dimensional Schr\"odinger equation
for the channel wave function $U_c(n)$ in the absence of all coupling terms
$v_{\lambda,c}$
\be\label{uceq}
-t_c U_c(n-1) + [V_c(n) + 2 t_c +E_c -E] U_c(n) - t_c U_c(n+1)  = 0\,\,\,  (1\le n \le N) \;,
\ee 
where $V_c(n)$ is the channel potential at site $n$.
The wave function $U_c(n)$ is assumed to be real and can be written in terms of 
 incoming and outgoing wave functions, $I_c(n)$ and $I_c^*(n)$, respectively,
\be\label{ucdef}
U_c(n) = i[I_c(n) - I_c^*(n)] \;.
\ee
 At the upper mesh points where the potential $V_c(n)$ can be ignored, the incoming wave $I_c(n)$ has the following asymptotic form
\be\label{asympIc}
I_c(n) = e^{-i\delta_c} e^{-i\kappa_c n} \;,
\ee
where $\delta_c$ is the phase shift for scattering in a potential $V_c$. 

At $n=1$,  the real wave function $U_c$  satisfies the Hamiltonian equation
\be\label{uceq2}
[2t_c\cos\kappa_c + V_c(1)]U_c(1) - t_c U_c(2) = 0 \;,
\ee
where we have used $U_c(0)=0$ and the dispersion relation (\ref{k_c}).

After including the interaction $v_{\lambda c}$  with the internal wave function amplitudes, the channel wave function acquires 
a different mixture of incoming and outgoing waves
\be\label{phicdef}
\varphi_c(n) = a_c^{(-)} I_c(n) - a_c^{(+)} I_c^*(n) \;.
\ee
Eliminating the internal state amplitudes, the Hamiltonian equation acting at site $n=1$ has the form (\ref{Hcc'}) but with the additional contribution of the channel potential
\be\label{phiceq}
[2 t_c\cos\kappa_c + V_c(1)]\varphi_c(1) - t_c\varphi_c(2) + \sum_{c^\prime =1}^{N_c} \sum_{\lambda = 1}^{N_i} \frac{v_{\lambda c} v_{\lambda c^\prime}}{E - E_\lambda} \varphi_{c^\prime}(1) = 0 \;.
\ee 
Multiplying Eq.~(\ref{uceq2}) by  $\varphi_c(1)$,  and 
subtracting it from Eq.~(\ref{phiceq}) multiplied by $U_c(1)$, we obtain
\be\label{gbeq1}
-t_c\left[\varphi_c(2)U_c(1) - U_c(2)\varphi_c(1) \right] + \sum_{c^\prime , \lambda}U_c(1) \frac{v_{\lambda c} v_{\lambda c^\prime}}{E - E_\lambda} \varphi_{c^\prime}(1) = 0 \;.
\ee
Inserting Eqs.~(\ref{ucdef}) and (\ref{phicdef}) in Eq.~(\ref{gbeq1}) and simplifying, yields
\be\label{gbeq2}
-it_c\mathcal{W}_c(a_c^{(-)} - a_c^{(+)}) + \sum_{c^\prime , \lambda}U_c(1) \frac{v_{\lambda c} v_{\lambda c^\prime}}{E - E_\lambda}(a_{c^\prime}^{(-)}I_{c^\prime}(1) - a_{c^\prime}^{(+)}I_{c^\prime}^*(1)) = 0  \;,
\ee
where
\be\label{Wcdef}
\mathcal{W}_c = I_c(1)I_c^*(2) - I_c^*(1)I_c(2) \;.
\ee
In analogy with the Wronskian of a second-order differential operator,
${\mathcal{W}_c(n)=  I_c(n)I_c^*(n+1) - I_c^*(n)I_c(n+1)}$ is independent of the mesh position $n$ and thus can be
evaluated in the asymptotic regime to give
\be
\mathcal{W}_c = 2i \sin \kappa_c \;.
\ee
The current $J^{c\,(\mp)}_{n,n+1}$ is given an expression similar to Eq.~(\ref{current}) but with the wave functions  $a_c^{(-)} I_c$  and $a_c^{(+)} I^*_c$. It is proportional to $\mathcal{W}_c$ and is thus independent of the mesh position, leading to the same result  (\ref{model_curr}) as in the case without potential interactions.
As in the main text, we introduce the flux-normalized variables
\be\label{bcdef}
b_c^{(\pm)} = a_c^{(\pm)}/d_c \;.
\ee
Inserting Eq.~(\ref{bcdef}) into Eq.~(\ref{gbeq2}), multiplying both sides by $d_c$, and separating the terms in that are proportional to $b_c^{(-)}$ and $b_c^{(+)}$ yields
\be\label{gbeq4}
\sum_{c^\prime} \bigg(\delta_{cc^\prime} + \sum_\lambda U_c(1) \frac{d_c v_{\lambda c} v_{\lambda c^\prime} d_{c^\prime}}{E - E_\lambda}I_{c^\prime}(1)\bigg) b_{c^\prime}^{(-)} = \sum_{c^\prime} \bigg(\delta_{cc^\prime} + \sum_\lambda U_c(1) \frac{d_c v_{\lambda c} v_{\lambda c^\prime} d_{c^\prime}}{E - E_\lambda}I^*_{c^\prime}(1)\bigg) b_{c^\prime}^{(+)} \;.
\ee
Thus, the $S$ matrix is given by $S = (D^*)^{-1} D$, where the $D$ matrix is now defined by
\be\label{Ddef2}
D_{cc^\prime} = \delta_{cc^\prime} + \sum_\lambda U_c(1) \frac{d_c v_{\lambda c} v_{\lambda c^\prime} d_{c^\prime}}{E - E_\lambda}I_{c^\prime}(1) \;.
\ee
The corresponding $S$ matrix can be shown to be unitary and symmetric. 

Finally, in the absence of a channel potential, $I_c(1)=e^{-i\kappa_c}$, $U_c(1)= 2\sin \kappa_c$, and the $D$ matrix in Eq.~(\ref{Ddef2}) reduces  to Eq.~(\ref{Ddef}).

\end{document}